\documentclass[prb,amsmath,amssymb]{revtex4}

\usepackage{graphicx}

\begin{document}
%
%
\title{Relating on-shell and off-shell formalism in perturbative quantum field theory}

\author{Christian Brouder}
\email[]{christian.brouder@impmc.jussieu.fr}
\affiliation{Institut de Min\'eralogie et de Physique des Milieux Condens\'es,
CNRS UMR7590, Universit\'es Paris 6 et 7, IPGP, 4 place Jussieu,
F-75252 Paris Cedex 05, France.}
\author{Michael D\"utsch}
\email[]{duetsch@physik.unizh.ch}
\affiliation{Max Planck Institute for Mathematics in the Sciences,
Inselstrasse 22, D-04103 Leipzig, Germany.}
\date{\today}
\begin{abstract}
In the on-shell formalism (mostly used in perturbative quantum field theory)
the entries of the time ordered product $T$ are on-shell fields (i.e. the
basic fields satisfy the free field equations). With that,
(multi)linearity of $T$ is incompatible with the Action Ward identity. This can 
be circumvented by using the off-shell formalism in which the entries of $T$ 
are off-shell fields. To relate on- and off-shell formalism correctly, 
a map $\sigma$ from on-shell fields to off-shell fields was introduced 
axiomatically by D{\"{u}}tsch and Fredenhagen\cite{Dutsch03}. 
In that paper it is shown that, in the case of one real scalar field in 
$N=4$ dimensional Minkowski space, these axioms have a unique 
solution. However, this solution is given there only recursively. 
We solve this recurrence relation and give a fully explicit expression for 
$\sigma$ in the cases of the scalar, Dirac and gauge fields for arbitrary values of 
the dimension $N$.
\end{abstract}
\maketitle
\section{Introduction}
The time-ordered product was introduced by Dyson \cite{Dyson}
to order products of fields so that a field $\varphi(x)$
is on the left of $\varphi(y)$ if $x^0>y^0$.
For example,
\begin{eqnarray}
T\big(\varphi(x),\varphi(y)\big) &=&
\theta(x^0-y^0)\varphi(x)\varphi(y)
+\theta(y^0-x^0) \varphi(y)\varphi(x).\label{Tnaive}
\end{eqnarray}
Applying Wick's theorem to the products of fields,
$T(\varphi(x),\varphi(y))$ becomes the sum of the normal
product ${:}\varphi(x)\varphi(y){:}$ and
the pointwise products of distributions
\begin{eqnarray*}
\theta(z^0)\,\Delta_m^+(z)+\theta(-z^0)\,\Delta_m^+(-z)\ ,\quad z\equiv x-y\ ,
\end{eqnarray*}
where $\Delta_m^+(x-y)\equiv <0|\varphi(x)\varphi(y)|0>$ is the 
two-point function to the mass $m$. This product is a priori defined only for $z\not= 0$, 
i.e.~in $\mathcal{D}'(\mathbf{R}^4\setminus\{ 0\})$, where 
$\mathcal{D}'(\Omega)$ is the space of distributions (i.e. the dual of 
of $\mathcal{D}(\Omega)$, the space of smooth functions compactly supported on the open subset $\Omega$
of $\mathbf{R}^N$). In the present case the extension of this product to 
$\mathcal{D}'(\mathbf{R}^4)$ can be done by continuity giving the Feynman propagator
$\Delta^F_m(z)$. This extension is unique if one requires that the extended distribution 
is not ``more singular'' at the origin $z=0$ than the non-extended one. This can be formulated 
rigorously in terms of Steinmann's scaling degree \cite{Steinmann0,Brunetti2}.

Applying the time ordering prescription (\ref{Tnaive}) to 
$T\big(\partial^\mu\varphi,\partial^\nu\varphi\big)$, the problematic term reads
\begin{eqnarray*}
\theta(z^0)\,\partial^\mu\partial^\nu\Delta_m^+(z)+\theta(-z^0)\,
\partial^\mu\partial^\nu\Delta_m^+(-z)\in\mathcal{D}'(\mathbf{R}^4\setminus\{ 0\})
\end{eqnarray*}
(a global minus-sign is omitted).
However, the extension to $\mathcal{D}'(\mathbf{R}^4)$ is no longer unique, even if we
require that it is Lorentz covariant and does not increase the degree of singularity at $z=0$.
The general form of the extensions satisfying these two conditions is
\begin{eqnarray*}
\partial^\mu\partial^\nu\Delta^F_m(z)+C\,g^{\mu\nu}\,\delta(z)\ ,
\end{eqnarray*}
with an arbitrary constant $C\in\mathbf{C}$.
Apart from very few exceptions \cite{scalQED,YM1,Scharf2} the choice
$C=0$
is used in the physical literature, see e.g.~the treatment of scalar QED in Sect.~6-1-4
of Ref.~\onlinecite{Itzykson}.
This choice is distinguished by the property
\begin{eqnarray}
T\big(\partial^\mu\varphi(x),\partial^\nu\varphi(y)\big) &=&
\partial^\mu_x \partial^\nu_y
T\big(\varphi(x),\varphi(y)\big).
\label{Tcovdd}
\end{eqnarray}

The study of the proper definition of 
$T\big(\partial^\mu\varphi,\partial^\nu\varphi\big)$
started in the late fourties
\cite{Matthews50,Matthews50E,Rohrlich,Nishijima},
where it is sometimes called the $T^*$-product.
An elementary exposition of this question can be
found in several textbooks
(see Refs.~\onlinecite{Itzykson} p.~284, \onlinecite{Sterman} p.~114,
\onlinecite{SchweberRQFT} p.~482).

For time-ordered products involving derivatives of higher order,
it would be tempting to generalize equation (\ref{Tcovdd}) and
to define, for example,
\begin{eqnarray*}
T\big(\partial^\mu\partial^\nu\varphi(x),\varphi(y)\big) &=&
\partial^\mu_x\partial^\nu_x T\big(\varphi(x),\varphi(y)\big)\ .\label{Tdd}
\end{eqnarray*}
However, this definition would not be compatible with
(multi)linearity of the map $T$ and
the fact that the on-shell field $\varphi(x)$ satisfies
the equation of motion $(\Box+m^2) \varphi(x)=0\,$,
because
\begin{eqnarray*}
T\big((\Box+m^2)\varphi(x),\varphi(y)\big) =
T\big(0,\varphi(y)\big) = 0\ ,
\end{eqnarray*}
whereas
\begin{eqnarray*}
(\Box_x+m^2) T\big(\varphi(x),\varphi(y)\big) &=&
(\Box_x+m^2)\Delta_m^F(x-y)\\
&=& -i\delta(x-y)\not=0.
\end{eqnarray*}

This interplay between the derivative of time-ordered products
and the equation of motion made the definition of a covariant
time ordered product of fields with higher order derivatives a 
longstanding problem of perturbative quantum field
theory (see e.g.  Sect. V of Ref.~\onlinecite{Lam72}). 
This problem has been solved over the years on a case by case 
basis, for example in perturbative QCD 
(see e.g. Ref.~\onlinecite{Pascual})
and in non-commutative Abelian gauge theory \cite{RimYee}.
A more specific example is the elimination of Schwinger terms.
\cite{GrossJackiw}

Finally, a recursive construction of the covariant time-ordered product 
of fields with derivatives of arbitrary high orders was recently 
given\cite{Dutsch03,Dutsch04}. 
In the present paper, we elaborate
on the results of these authors by solving their recursive equation
and giving a fully explicit definition of the time-ordered product
of on-shell fields.

We now describe very briefly the general construction.
The first idea is that, when the field
is not assumed to satisfy the equation of motion 
(off-shell field), then one can renormalize the corresponding time ordered
product $T_\mathrm{off}$ so that
the spacetime derivatives commute with the map $T_\mathrm{off}$, e.g.
(\ref{Tcovdd}) and (\ref{Tdd}) hold true for $T_\mathrm{off}$.
This is called the Action Ward Identity
(AWI), as we shall see. 
To describe the second idea, we first set up some notation.
If $N$ is the dimension of spacetime and 
$a=(\alpha_0,\dots,\alpha_{N-1})$ is a N-tuple of nonnegative integers,
we denote by $\partial^a\varphi(x)$ the
partial derivative
\begin{eqnarray*}
\partial^a\varphi(x) &=&
\frac{\partial^{\alpha_0+\dots+\alpha_{N-1}}\varphi(x)}
     {\partial x_0^{\alpha_0}\dots \partial x_{N-1}^{\alpha_{N-1}}}\ .
\end{eqnarray*}
The second idea is to define a map $\sigma$ that 
associates, to any derivative of the
on-shell field $\partial^a\varphi$, a sum of
derivatives of the off-shell field. With that, the time ordered product 
$T$ of on-shell
fields is defined in terms of the time ordered product $T_\mathrm{off}$
of off-shell fields by setting
\begin{eqnarray}
T\Bigl(\prod_j(\partial^{a_j}\varphi(x)),...\Bigr):=
T_\mathrm{off}\Bigl(\prod_j(\sigma(\partial^{a_j}\varphi)(x)),...\Bigr)\ .
\label{Ton=Toff}
\end{eqnarray}
The r.h.s.~is constructed by using the AWI.
It is remarkable that, under very natural conditions,
the map $\sigma$ is essentially unique. The present paper
provides an explicit description of this map.

After this introduction, we describe precisely what we mean
by on-shell and off-shell fields and we present the 
axiomatic definition of the map $\sigma$. Then, we
transform this axiomatic definition into an explicit
description for the scalar, Dirac and gauge fields.

\section{On- and off-shell fields}
In the present paper, we consider commutative algebras of fields
with respect to a pointwise product (by following 
Refs.~\onlinecite{Dutsch03} and \onlinecite{Dutsch04}).
For on-shell fields, this corresponds to the usual quantum fields 
(which are distributions with values in the Fock space operators)
equipped with the normal product: $(\partial^a\varphi(x)\,,\,\partial^b\varphi(x))\mapsto$
$:\partial^a\varphi(x)\,\partial^b\varphi(x):\,$, or more generally
\begin{eqnarray}
\Bigl(:\prod_j(\partial^{a_j}\varphi(x)):\,,\,:\prod_k(\partial^{b_k}\varphi(x)):\Bigr)
\mapsto \,:\prod_j(\partial^{a_j}\varphi(x))\,\prod_k(\partial^{b_k}\varphi(x)):
\end{eqnarray}
The algebra generated by the set $\{\partial^a\varphi\,|\,a\in {\bf N}_0^N\}$
with respect to this product is called the algebra of local Wick 
polynomials $\mathcal{W}_\mathrm{loc}$.

We first introduce the commutative algebra $\mathcal{P}$ of
off-shell fields. An off-shell field $\varphi(x)$ is defined,
similarly to the classical field \cite{Dutsch,DFFields,Dutsch03},
as an evaluation functional on the classical configuration space
$\mathcal{C}^\infty({\bf R}^N)$: it acts on an $h\in \mathcal{C}^\infty({\bf R}^N)$ by
$\big(\varphi(x)\big)(h)=h(x)$. Space-time derivatives of fields act on $h$ by
$(\partial^a\varphi)(x)(h)=\partial^a h(x)$ and the product
of fields is defined by
\begin{eqnarray*}
\big(\partial^{a_1}\varphi(x) \dots \partial^{a_n}\varphi(x)\big)(h)
&=&
\partial^{a_1} h(x) \dots \partial^{a_n} h(x).
\end{eqnarray*}
Or, smeared with a test function $f\in\mathcal{D}(\mathbf{R}^N)$, this equation reads
\begin{eqnarray*}
\big(\prod_j(\partial^{a_j}\varphi)(f)\big)(h)&\equiv&
\big(\int dx\,\prod_j(\partial^{a_j}\varphi(x))\,f(x)\big)(h)\nonumber\\
&=& \int dx\,\prod_j(\partial^{a_j}h(x))\,f(x)\ .
\end{eqnarray*}
The commutative algebra $\mathcal{P}$ is the algebra 
of polynomials in the off-shell basic field $\varphi$
and its partial derivatives, that is
$\mathcal{P}:= \bigvee\{\partial^a\varphi\,|\,a\in{\bf N}_0^N\}$,
where $\bigvee M$ denotes the algebra generated by the elements of
the set $M$.

To go from the off-shell to the on-shell fields, we first
denote by $\mathcal{J}$ the ideal in $\mathcal{P}$ 
generated from the free field equation,
\begin{equation}
        \mathcal{J}=\{\sum_{a \in {\bf N}_{0}^{N}}
        B_{a}\partial^{a}(\square +m^2)\varphi\,|\, 
        B_{a}\in\mathcal{P}\} \ .
        \label{eq:ideal of free fields}
\end{equation}
The quotient of the algebra $\mathcal{P}$ by the ideal
$\mathcal{J}$ is denoted by $\mathcal{P}_{0}$.
Let $\pi:\mathcal{P}\to\mathcal{P}_{0}\,,\,\,
\pi(A)=A+\mathcal{J}$ 
be the canonical surjection. Obviously, $\pi$ is an algebra homomorphism, i.e.
it is linear and commutes
with multiplication, $\pi(AA')=\pi(A) \pi(A') \quad (\forall A, A'\in\mathcal{P}) $.
Derivatives in $\mathcal{P}_{0}$ are defined as follows:
for $A\in\mathcal{P}_{0}$ choose a 
$B\in\mathcal{P}$ such that $\pi(B)=A$. Then,
$\partial^\mu A:=\pi \partial^\mu B$ is well defined, since
$A=\pi(B_1)=\pi(B_2)$ implies $(\partial^\mu B_1-\partial^\mu B_2)\in\partial^\mu
\mathcal{J}\subset\mathcal{J}$.
In particular, $(\square +m^2)\pi\varphi=\pi(\square +m^2)\varphi=0$,
because $(\square +m^2)\varphi\in\mathcal{J}$. In other
words, the element $\pi\varphi$ of $\mathcal{P}_{0}$ satisfies the
free field equation. Besides,  $\mathcal{P}_{0}$
can be identified with the algebra of local Wick polynomials
$\mathcal{W}_\mathrm{loc}$, i.e. there is an algebra 
isomorphism $\mathcal{P}_{0}\rightarrow\mathcal{W}_\mathrm{loc}$
(cf. Sect.~5.1 of Ref.~\onlinecite{Dutsch}, or Ref.~\onlinecite{DFFields}).

\section{The time-ordered product}

We mentioned in the introduction that already the 
time-ordered product of derivatives of the basic field(s)
cannot be defined as the naive ordering of the
fields according to their time variable;
for the time ordered product of field polynomials $A(x)\,\,(A\in\mathcal{P}_0)$,
the problem is even much harder due to the appearance of loop diagrams.
We work with causal perturbation theory in which the time-ordered product is 
defined by a number of desirable properties. This
axiomatic approach to time-ordered product
has a long history: starting with
Stueckelberg \cite{StueckelbergR} and 
Bogoliubov \cite{Bogoliubov} it was worked out rigorously 
by Epstein and Glaser \cite{Epstein} and generalized 
to curved space times \cite{Brunetti2,Hollands2}.

We explain only the axioms that are
relevant to the present work. 
In quantum field theory textbooks \cite{Peskin,Itzykson}, 
the perturbative calculation of the S-matrix is made
by evaluating the time-ordered products
$\int d x_1\dots d x_n \, T(\mathcal{L}(x_1),\dots,\mathcal{L}(x_n))$,
where $\mathcal{L}\in \mathcal{W}_\mathrm{loc}$ is the Lagrangian.
To avoid infrared divergences, we replace the
interaction $I=\int d x\,\mathcal{L}(x)$ by the smeared interaction
$I=\mathcal{L}(g)=\int d x\, g(x) \mathcal{L}(x)$, where $g\in\mathcal{D}(\mathbf{R}^N)$.
The S-matrix is evaluated perturbatively by computing
$\int d x_1\dots d x_n \,g(x_1)\dots g(x_n)\,
  T(\mathcal{L}(x_1),\dots,\mathcal{L}(x_n))$ for all $n$.
The usual formulation is recovered by taking the adiabatic limit
$g\to 1$ in the end.

The on-shell time ordered product of $n$-th order, $T_n\equiv T$,
is a map from $\mathcal{P}_0^{\otimes n}$ (or $\mathcal{W}_\mathrm{loc}^{\otimes n}$)
into the space of distributions on $\mathcal{D}({\bf R}^{Nn})$ with values in the 
Fock space operators. Hence, it would be more rigorous to write
$T_n(\mathcal{L}_1\otimes\dots\otimes \mathcal{L}_n)(x_1,\dots ,x_n)$
instead of $T(\mathcal{L}_1(x_1),\dots,\mathcal{L}_n(x_n))$ (which we use for convenience)
or $T(\mathcal{L}_1(x_1)\dots\mathcal{L}_n(x_n))$ (which is mostly used in textbooks).
For the off-shell time ordered product of $n$-th order, $T_{\mathrm{off},n}\equiv 
T_\mathrm{off}$, the difference is that the domain is $\mathcal{P}^{\otimes n}$
(instead of $\mathcal{P}_0^{\otimes n}$); the axioms we are going to recall for $T$ are 
used to define also $T_\mathrm{off}$.\cite{chifn1}

A first property that is true for the naive time-ordered product (\ref{Tnaive})
and that we want to retain is symmetry of $T$:
\begin{eqnarray*}
T(\mathcal{L}_{\pi 1}(x_{\pi 1}),\dots,\mathcal{L}_{\pi n}(x_{\pi n}))=
T(\mathcal{L}_1(x_1),\dots,\mathcal{L}_n(x_n))
\end{eqnarray*}
for all permutations $\pi$.
A second desirable property is linearity of the map $T$:
\begin{eqnarray*} 
T((\mathcal{L}_1+\lambda \mathcal{L})(x_1),\mathcal{L}_2(x_2),\dots) =
T(\mathcal{L}_1(x_1),\mathcal{L}_2(x_2),\dots)
+\lambda\,T(\mathcal{L}(x_1),\mathcal{L}_2(x_2),\dots)\ ,
\end{eqnarray*}
where $\lambda\in\mathbf{C}$. The 'initial condition' determines the
time ordered product of first order; in the on-shell formalism
it reads $T_1(\mathcal{L}(x))=\mathcal{L}(x)$.
As indicated by the word 'causal',
the most striking defining property of $T$ is causality; however, we shall 
not use it explicitly and, hence, refer to the literature \cite{Bogoliubov,Epstein}.

\section{The Action Ward Identity}

The action Ward identity requires that the off-shell $S$-matrix 
depends only on the 
interaction $I=\int dx\, g(x)\, \mathcal{L}(x)$ with $g\in\mathcal{D}$, 
and not on the choice of a 
corresponding Lagrangian $\mathcal {L}$. 
This seemingly obvious requirement has
striking consequences\cite{Stora02,Storaprivcom,Stora06}. 
For example, it implies that 
$T_\mathrm{off}$ commutes with derivatives, as we show now.
For a Lagrangian
\begin{eqnarray*} 
g(x)\mathcal{L}(x)+\lambda\, \partial^\mu_x (f(x)\mathcal{L}_1(x))
\quad (f,g\in\mathcal{D}\,,\,\,\mathcal{L},\mathcal{L}_1\in\mathcal{P}) 
\end{eqnarray*} 
the AWI requires that the corresponding $S$-matrix is independent of $\lambda$.
Using linearity and symmetry of $T_\mathrm{off}$, this condition implies
\begin{eqnarray*} 
0&=&\frac{d}{d\lambda}\Big\vert_{\lambda=0}\int dx_1...dx_n
\sum_{l=0}^n \sum_{k=0}^l
\frac{\lambda^l\, n!}{k!(l-k)!(n-l)!}\,f(x_1)...f(x_k)\partial^\mu f(x_{k+1})...\partial^\mu f(x_l)\\
&&g(x_{l+1})...g(x_n)\,T_\mathrm{off}(\partial^\mu \mathcal{L}_1(x_1),...,\partial^\mu \mathcal{L}_1(x_k),
\mathcal{L}_1(x_{k+1}),...,\mathcal{L}_1(x_{l}),\mathcal{L}(x_{l+1}),...,\mathcal{L}(x_n))\\
&=& n \int dx_1...dx_n\,\Bigl(f(x_1)g(x_2)...g(x_n)\,T_\mathrm{off}(\partial^\mu \mathcal{L}_1(x_1),
\mathcal{L}(x_{2}),...,\mathcal{L}(x_n))\\
&&+(\partial^\mu f)(x_1)g(x_2)...g(x_n)\,T_\mathrm{off}(\mathcal{L}_1(x_1), 
\mathcal{L}(x_{2}),...,\mathcal{L}(x_n))\Bigr) 
\end{eqnarray*} 
for all $f,g\in\mathcal{D}$ and all $\mathcal{L},\mathcal{L}_1\in\mathcal{P}$.
We integrate $\partial^\mu f$ by parts and use the fact that
the equality is true for any $f$ and $g$ to obtain
\begin{eqnarray*} 
\partial^\mu_x\,T_\mathrm{off}(\mathcal{L}_1(x),...)=T_\mathrm{off}(\partial^\mu \mathcal{L}_1(x),...)\ , 
\quad\forall \mathcal{L}_1\in\mathcal{P}\ . 
\end{eqnarray*} 
This is a renormalization condition. 

In Ref.~\onlinecite{Dutsch04} a time ordered product $T_{\rm off}$ 
is constructed which satisfies permutation symmetry,
linearity, causality, the AWI, Poincar\'e 
covariance and further renormalization conditions. 
The time-ordered product $T$ in the on-shell 
formalism is then given by (\ref{Ton=Toff}), which we rewrite in the form
\begin{eqnarray}
T(\mathcal{L}_1(x_1),...,\mathcal{L}_n(x_n))=
T_{\rm off}(\sigma(\mathcal{L}_1)(x_1),...,\sigma(\mathcal{L}_n)(x_n)),
\label{T_on-T_off}
\end{eqnarray}
where $\mathcal{L}_1,\dots,\mathcal{L}_n$ are in 
$\mathcal{P}_0$ and $\sigma:\mathcal{P}_0\rightarrow\mathcal{P}$ is an algebra 
homomorphism which chooses a representative $\sigma(A+\mathcal{J})
\in A+\mathcal{J}$ of the equivalence class $A+\mathcal{J}$, that is
$\pi\circ\sigma=\mathrm{id}$ \cite{Dutsch03}.
To relate on- and off-shell formalisms correctly, 
$\sigma$ must satisfy certain conditions that will be described
in the following section.

\medskip
\noindent{\it Remark (1)}:
Since $\sigma$ cannot be surjective, the set $\{T(A_1,\dots,A_n)|A_j$ 
runs through all on-shell fields$\,\}$
is significantly smaller than  $\{T_{\rm off}(B_1,\dots,B_n)|B_j$ 
runs through all off-shell fields$\,\}$.
This can be understood as
the reason why, in the usual on-shell formalism (i.e.~with the former set), 
it is impossible to formulate all Ward identities which one wants to hold, 
in particular the Master Ward Identity (MWI) 
\cite{Dutsch02,chifn2}.
To overcome this shortcoming an improved version of the on-shell formalism 
is introduced in Ref.~\onlinecite{Dutsch02}: the domain of $T$ is enlarged by 
introducing an ``external derivative''. The relation of this improved 
on-shell formalism to the off-shell formalism is clarified in Sect.~4 
of Ref.~\onlinecite{Dutsch03}. For the purpose of this paper 
it is not necessary to introduce the external derivative and, hence, 
we disregard it.
\section{The map $\sigma$ from on-shell fields to 
off-shell fields}
For a given on-shell element $A\in\mathcal{P}_0$,
there are generally many off-shell elements $B\in\mathcal{P}$
such that $\pi(B)=A$, that is the condition  $\pi\circ \sigma=\mathrm{id}$
leaves a large freedom for $\sigma$.
Under a few further conditions, which are motivated by (\ref{T_on-T_off}),
the map $\sigma$ is unique for the models of a real scalar field and a 
Dirac field, as we will see.

For {\it one real scalar field}
the map $\sigma :\mathcal{P}_0\rightarrow\mathcal{P}$ is defined 
by the following axioms \cite{Dutsch03}:
\begin{enumerate}
    
     \item [(i)]  $\pi\circ \sigma=\mathrm{id}$.

     \item [(ii)] $\sigma$ is an algebra homomorphism.
     
     \item [(iii)] The Lorentz transformations commute with $\sigma\pi$.

     \item [(iv)]  $\sigma\pi(\mathcal{P}_1)\subset\mathcal{P}_1$, where 
$\mathcal{P}_1\subset\mathcal{P}$
is the subspace of fields linear in $\varphi$ and its partial derivatives.

     \item [(v)] $\sigma\pi$ does not increase the mass dimension of the fields. 

\end{enumerate}
It immediately follows $\sigma\pi(\varphi)=\varphi$.\cite{chifn3}
In Ref.~\onlinecite{Dutsch03} it is shown that in $N=4$ 
dimensional Minkowski space these axioms have a unique solution. 
However, this solution is given there only recursively.
 
The aim of this paper is to solve this recurrence relation and to give a fully explicit 
expression for $\sigma$ in $N$-dimensions, $N$ arbitrary, and to derive the corresponding 
result for the Dirac and gauge fields. 

For this purpose it is convenient to introduce another map $\chi$: let
$V$ be the real vector space generated by 
the partial derivatives $\partial_{0},
\dots, \partial_{N-1}$, we denote by
$S(V)$ the vector space of real polynomials 
in the variables $\partial_{0},\dots, \partial_{N-1}$.
Axiom (iv) is equivalent to the condition that there exists a map $\chi\>:\>S(V)\rightarrow S(V)$ such that
\begin{eqnarray}
\sigma\pi(u\,\varphi)=\chi(u)\,\varphi\ ,\quad\forall u\in S(V)\ .\label{sigma-chi}
\end{eqnarray}
$\chi$ is linear because $\sigma\pi$ has this property.
Because of (ii), the map $\sigma$ is completely determined by $\chi$.

%

We claim that $\chi$ must satisfy the following conditions:
\begin{enumerate}
\item [(a)] $\chi$ is linear and symmetric: 
$\chi(\partial_{\mu_{\tau(1)}}\dots\partial_{\mu_{\tau(n)}})
=\chi(\partial_{\mu_1}\dots\partial_{\mu_n})$ for all permutations $\tau\in \mathcal{S}_n$.
\item [(b)] $\chi\big((\Box+m^2) u\big)=0$ for all $u\in S(V)$.
\item [(c)]
  For any $u=\partial_{\mu_1} \dots \partial_{\mu_n}$,
  we want that $\chi(u)$ transforms under Lorentz transformation
  as $u$. 
\item [(d)] $\chi$ does not increase the degree of the derivatives,
i.e. $\chi(\partial_{\mu_1}\dots\partial_{\mu_n})$ is a polynomial of degree $\leq n$.
\item [(e)] On monomials $\chi$ takes the form
\begin{eqnarray*}
\chi(\partial_{\mu_1}\dots\partial_{\mu_n})&=&\partial_{\mu_1}\dots\partial_{\mu_n}
+\sum_{i<j}g_{\mu_i\mu_j}\,p_{\mu_1\dots\hat i\dots\hat j\dots\mu_n}\ ,
\end{eqnarray*}
where the carets $\hat i$ and $\hat j$ mean that the indices $\mu_i$  and $\mu_j$ are omitted and
where $p_{\mu_1\dots\mu_{n-2}}\in S(V)$ is symmetric in $\mu_1,\dots,\mu_{n-2}$ and Lorentz covariant.
In particular
\begin{eqnarray*}
\chi(1)&=&1\ ,\quad\chi(\partial_\mu)=\partial_\mu\ .
\end{eqnarray*}
\end{enumerate}

To prove this claim we first note that, obviously, (a) follows from
(\ref{sigma-chi}), (c) from (iii) and (d) from (v). As pointed out 
in Ref.~\onlinecite{Dutsch03}, (i) implies 
$\mathrm{ker}\,\sigma\pi = \mathrm{ker}\,\pi=\mathcal{J}$, 
this gives (b). To derive (e) we first note that (i) and (iv) imply 
$\sigma\pi(u\varphi)-u\varphi\in\mathrm{ker}\,\pi\cap\mathcal{P}_1 =
\mathcal{J}\cap\mathcal{P}_1$. 
Taking additionally (a), (c) and (d) into account we conclude that $\chi$ is of the form
\begin{eqnarray}
\chi(\partial_{\mu_1}\dots\partial_{\mu_n})&=&
\partial_{\mu_1}\dots\partial_{\mu_n} +\sum_{l=0}^{n-2}
c_{\mu_1\dots\mu_n}^{\nu_1\dots\nu_l}\,\partial_{\nu_1}\dots\partial_{\nu_l}
(\Box+m^2)\ ,\label{chi=}
\end{eqnarray}
where $c_{\mu_1\dots\mu_n}^{\nu_1\dots\nu_l}\in {\bf R}$ is a constant Lorentz 
tensor which is symmetric in $\mu_1\dots\mu_n$ (cf. formula (83) in 
Ref.~\onlinecite{Dutsch03}). Since the sum over $l$ runs only up to $(n-2)$, 
(e) follows.

The conditions (a)-(e) are not only necessary for $\sigma$, together 
with (ii) and (\ref{sigma-chi}) they are also {\it sufficient}. This is obvious
except for (i), which can be expressed by
$0=\pi\circ\sigma\circ\pi(u\varphi)-\pi(u\varphi)=\pi(\chi(u)-u)\varphi$.
That is we have to show that (a)-(e) imply that
$\chi(u)-u$ is of the form
\begin{eqnarray*}
\chi(u)-u=\sum_a c_a\,\partial^a(\Box+m^2)\ ,\quad c_a\in\mathbf{R}\ ,\,\,\forall u\in S(V)\ .
\end{eqnarray*}
But, as worked out in Sect.~3, the solution $\chi$ of (a)-(e) is indeed of this form. Or, 
one can also argue as follows: as we will see,
the set of solutions of (a)-(e) is not bigger than for (i)-(v),
it has also precisely one element. Hence, the unique $\chi$ solving (a)-(e)
yields the unique $\sigma$ solving (i)-(v).
\medskip

The model of {\it one complex scalar field} $\phi$ can be viewed
as the model of two real scalar fields given by the real and imaginary part of $\phi$;
we write $\phi=\varphi_1+i\,\varphi_2$ with $\varphi_1=\varphi_1^*$ and $\varphi_2=\varphi_2^*$.
$\mathcal{P}$, $\mathcal{P}_1$ and $\mathcal{J}$ are modified: $\mathcal{P}$ is the complex 
$\star$-algebra generated by $\partial^a\phi$ and $\partial^a\phi^*$ ($a\in{\bf N}_0^N$). 
$\mathcal{P}_1$ is the subspace of fields linear in $\phi$ and $\phi^*$ and their partial derivatives, and
$\mathcal{J}$ is the ideal in $\mathcal{P}$ generated by $(\Box+m^2)\phi$ and $(\Box+m^2)\phi^*$. With that 
the axioms (i)-(v) are well-defined. As mentioned in Ref.~\onlinecite{Dutsch03}, the axiom
\begin{enumerate}
\item [(iia)] $\sigma\pi (A^*)=\sigma\pi (A)^*\ ,\,\forall A\in\mathcal{P}\ $,
\end{enumerate}
has to be added. It follows $\sigma\pi(\phi)=\phi$ and $\sigma\pi(\phi^*)=
\phi^*$.\cite{chifn4}
The most obvious solution\cite{chifn5} is obtained from the
(unique) map $\sigma_\mathrm{real}$ of the real scalar field (treated above)
by setting
\begin{eqnarray*}
\sigma\pi (u\phi)&=&\sigma_\mathrm{real}\pi (u\varphi_1)+i\,
\sigma_\mathrm{real}\pi (u\varphi_2)\\ 
&=& \chi(u)\,\phi\,,\quad\forall u\in S(V)\ ,
\end{eqnarray*}
where $\chi$ is given in terms of $\sigma_\mathrm{real}$ by (\ref{sigma-chi}) and, 
as above, the elements of $S(V)$ are real. It follows $\sigma\pi (u\,\phi^*)=
\chi(u)\,\phi^*$ with the same map $\chi$. Hence, the unique $\chi$ solving (a)-(e)
yields also a map $\sigma$ for the complex scalar field.
\medskip

We turn to the model of {\it one Dirac field} $\psi\in\mathbf{C}^{f(N)}$, where $f(N)$ is the 
size of the $\gamma$-matrices, which is $f(N)=N$ if 
the spacetime dimension $N$ is even and $f(N)=N-1$ if $N$ is odd. The $\gamma$-matrices
satisfy the relations $\gamma_\mu\gamma_\nu+\gamma_\nu\gamma_\mu=2 g_{\mu\nu}$
and $\gamma_0^+=\gamma_0\ ,\ \gamma_j^+=-\gamma_j$. The field ``algebra'' $\mathcal{P}$
is not an algebra, it is a complex vector space with a product which is only partially defined.
The vector space reads
\begin{eqnarray*}
\mathcal{P}=\mathcal{P}_\mathrm{scalar}\oplus \mathcal{P}_\mathrm{spinor}\oplus 
\mathcal{P}_\mathrm{spinor}^+\oplus \mathcal{P}_\mathrm{matrix}\ ,
\end{eqnarray*}
where
\begin{eqnarray*}
\mathcal{P}_\mathrm{scalar}&=&\bigvee\{\partial^a\psi^+\gamma_{\mu_1}\dots\gamma_{\mu_k}\partial^b\psi\,|\,
a,b\in {\bf N}_0^N\,,\,k\in {\bf N}_0\}=\mathcal{P}_\mathrm{scalar}^+\ ,\\
\mathcal{P}_\mathrm{spinor}&=&\mathcal{P}_\mathrm{scalar}\cdot \Bigl[\{
\gamma_{\mu_1}\dots\gamma_{\mu_k}\partial^a\psi\,|\, a\in {\bf N}_0^N\,,\,k\in {\bf N}_0\}\Bigr]\ ,\\
\mathcal{P}_\mathrm{matrix}&=&\mathcal{P}_\mathrm{scalar}\cdot \Bigl[\{\gamma_{\mu_1}\dots\gamma_{\mu_k}\,,\,
\gamma_{\mu_1}\dots\gamma_{\mu_k}\partial^a\psi\,\partial^b\psi^+\gamma_{\nu_1}\dots\gamma_{\nu_l}\,|\,
a,b\in {\bf N}_0^N\,,\,k,l\in {\bf N}_0\}\Bigr]=\mathcal{P}_\mathrm{matrix}^+\ .
\end{eqnarray*}
($\bigvee$ denotes the generated algebra and $[\quad ]$ the linear span.)
We point out that $A\equiv (A_1,A_2,A_3,A_4)\in\mathcal{P}$ has four components which belong to different spaces.
A $\star$-operation is given by complex conjugation and transposition:
\begin{eqnarray*}
\mathcal{P}\ni A=(A_1,A_2,A_3,A_4)\mapsto A^+=(A_1^+,A_3^+,A_2^+,A_4^+)\in\mathcal{P}\ ,
\end{eqnarray*}
where $A_j^+\equiv A_j^{*T}$.
The product is matrix multiplication which is defined only on the following set $\mathcal{M}\subset
\mathcal{P}\times\mathcal{P}$ (which is not a vector space):
\begin{eqnarray*}
\mathcal{M}&=&(\mathcal{P}_\mathrm{scalar}\times\mathcal{P})\cup
(\mathcal{P}\times\mathcal{P}_\mathrm{scalar})
\cup(\mathcal{P}_\mathrm{spinor}^+\times\mathcal{P}_\mathrm{spinor})\cup
(\mathcal{P}_\mathrm{spinor}\times\mathcal{P}_\mathrm{spinor}^+)\\
&&\cup(\mathcal{P}_\mathrm{matrix}\times\mathcal{P}_\mathrm{spinor})\cup
(\mathcal{P}_\mathrm{spinor}^+\times\mathcal{P}_\mathrm{matrix})
\cup(\mathcal{P}_\mathrm{matrix}\times\mathcal{P}_\mathrm{matrix})\ .
\end{eqnarray*}
 $\mathcal{J}\subset\mathcal{P}$ is the subspace of fields which vanishes 
modulo the Dirac equation and
the adjoint Dirac equation.  $\mathcal{J}$ can be written as
\begin{eqnarray*}
\mathcal{J}=\mathcal{J}_\mathrm{scalar}\oplus \mathcal{J}_\mathrm{spinor}\oplus 
\mathcal{J}_\mathrm{spinor}^+\oplus \mathcal{J}_\mathrm{matrix}=\mathcal{J}^+\ ,
\end{eqnarray*}
where e.g.
\begin{eqnarray*}
\mathcal{J}_\mathrm{matrix}&=&\{C\cdot M\cdot (\partial^{a}(i\,\gamma\cdot\partial-m)\psi)\cdot S^+\,,\,
C\cdot S\cdot (\partial^{a}\psi^+(i\,\overleftarrow{\partial}\cdot\gamma^+ +m))\cdot M\,|\,\\
&&C\in \mathcal{P}_\mathrm{scalar}\,,\,M\in \mathcal{P}_\mathrm{matrix}\,,\,
S\in \mathcal{P}_\mathrm{spinor}\,,\,a\in \mathbf{N}_0^N\}\ .
\end{eqnarray*}
Obviously the canonical projection 
$\pi\,:\,\mathcal{P}\rightarrow\mathcal{P}/\mathcal{J}\,,\,\pi(A)=A+\mathcal{J}$, satisfies
$\pi(A^+)=\pi(A)^+$. $\mathcal{J}$ is an ``ideal'' whenever matrix
multiplication is defined that is, if $(A,J_1)\in (\mathcal{P}\times\mathcal{J})\cap\mathcal{M}$ or
$(J_2,B)\in (\mathcal{J}\times\mathcal{P})\cap\mathcal{M}$, then it follows $A\cdot J_1\in\mathcal{J}$ or
$J_2\cdot B\in \mathcal{J}$, respectively. 
Therefore, we can define matrix multiplication for on-shell fields on the set
\begin{eqnarray*}
\{(\pi(A),\pi(B))\,|\,(A,B)\in\mathcal{M}\}\subset \mathcal{P}/\mathcal{J}\,\times\, \mathcal{P}/\mathcal{J}
\end{eqnarray*}
by setting $\pi(A)\cdot\pi(B):=\pi(A\cdot B)$. (We use here also that
$(A,B)\in\mathcal{M}$ implies that $A$ and $B$ are of a very restricted form, namely 
three components of $A$ and three components of $B$ vanish.)

To define the choice $\sigma\,:\,\mathcal{P}/\mathcal{J}\rightarrow\mathcal{P}$ 
of representatives, we use the axioms (i), 
(iii) and (v) as they are written above; (iia) reads now $\sigma\pi(A^+)=\sigma\pi(A)^+$
and, of course, $\sigma\pi$ acts trivially on $\gamma$-matrices: $\sigma\pi(\gamma_\mu)
=\gamma_\mu$.

Axiom (ii) requires now that $\sigma$ is a {\it linear} map which is diagonal
(i.e. $\sigma\pi(\mathcal{P}_a)\subset\mathcal{P}_a\,$, where $a$
stands for scalar, spinor or matrix) and which 
intertwines matrix multiplication:
$\sigma\pi(A\cdot B)=\sigma\pi(A)\cdot\sigma\pi(B)$ if $(A,B)\in\mathcal{M}$.

Axiom (iv) is now the condition that $\sigma\pi$ maps derivatives of $\psi$ into  a sum of derivatives 
of $\psi$. \cite{chifn6}
We formulate this analogously to (\ref{sigma-chi}): we modify $V$ and $S(V)$ to be
{\it complex} vector spaces, i.e. $S(V)$ is now the space of polynomials in
$\partial_0,\dots,\partial_{N-1}$ with complex coefficients. Axiom (iv) requires that there 
exists a map $\chi\,:\,S(V)\rightarrow S(V)^{f(N)\times f(N)}$ such that
\begin{eqnarray}
\sigma\pi(u\,\psi)=\chi(u)\,\psi\ ,\quad\forall u\in S(V)\ .\label{sigma-chi-Dirac}
\end{eqnarray}
Applying (iia) we obtain $\sigma\pi(u\,\psi^+)=\psi^+\,\overleftarrow{\chi(u^*)^+}$. From 
the axioms one easily derives
$\sigma\pi(\psi)=\psi$ and $\sigma\pi(\psi^+)=\psi^+$.

Analogously to the boson case the axioms for $\sigma$ imply that $\chi$ satisfies the conditions 
(a), (c), (d) and the modified conditions (b'), (e') which read:
\begin{enumerate}
\item [(b')] $i\gamma^\mu\,\chi(\partial_\mu\, u)-m\,\chi(u)=0\ ,\quad\forall u\in S(V)\ .$
\item [(e')] $\chi$ is of the form
\begin{eqnarray*}
\chi(\partial_{\mu_1}\dots\partial_{\mu_n})=\mathbf{1}_{f(N)\times f(N)}
\partial_{\mu_1}\dots\partial_{\mu_n}
+\sum_{i<j}g_{\mu_i\mu_j}\,p_{\mu_1\dots\hat i\dots\hat j\dots\mu_n}
+\sum_j\gamma_{\mu_j}\,q_{\mu_1\dots\hat j\dots\mu_n}\ ,
\end{eqnarray*}
where $p_{\mu_1\dots\mu_{n-2}}\in S(V)^{f(N)\times f(N)}$ and $q_{\mu_1\dots\mu_{n-1}}\in S(V)^{f(N)\times f(N)}$
are symmetric in $\mu_1,\mu_2,\dots$ and Lorentz covariant. $p_{\mu_1\dots\mu_{n-2}}$ and $q_{\mu_1\dots\mu_{n-1}}$
do not contain any $\gamma$-matrix $\gamma_{\mu_r}$ with an uncontracted Lorentz index $\mu_r\in
\{\mu_1,\dots,\mu_{n-2}(,\mu_{n-1})\}$, however they may contain $\gamma$-matrices with contracted Lorentz index,
e.g. $\gamma\cdot\partial\equiv\gamma_\nu\partial^\nu$.
In particular (e') requires $\chi(1)=1$.
\end{enumerate}
The derivation of (e') uses that, because of the anticommutation
properties of the $\gamma$-matrices, a symmetric Lorentz-covariant
tensor can contain at most one $\gamma$-matrix.
To show this, assume that a Lorentz-covariant tensor
contains a product of two $\gamma$-matrices
$\gamma_\mu\gamma_\nu$.
Then, the symmetry of the tensor implies that we
can replace $\gamma_\mu\gamma_\nu$
by $\gamma_\mu\gamma_\nu+\gamma_\nu\gamma_\mu=2 g_{\mu\nu}$
and the $\gamma$ matrices disappear.

As above, a solution $\chi$ of (a), (b'), (c), (d) and (e') determines 
uniquely a solution $\sigma$ of (i)-(v) (by using (\ref{sigma-chi-Dirac}) and (ii))
i.e.~(a)-(e') are also sufficient for $\sigma$. Again, it is not obvious that the so 
constructed $\sigma$ satifies (i). But, as it will turn out in Sect.~4, the (unique) 
solution $\chi$ of (a)-(e) satisfies $(\chi(u)-u)\psi\in\mathcal{J}\ ,\ \forall u\in S(V)$.
This implies $\pi\circ\sigma\circ\pi(u\psi)=\pi(u\psi)\ ,\ \forall u$,
and by means of (iia) we also get $\pi\circ\sigma\circ\pi(u\psi^+)=\pi(u\psi^+)\ ,\ \forall u$.

\medskip
\noindent{\it Remarks}: (2) For time ordered products with a factor $u\pi\varphi$ (real scalar field),
with $u\in S(V)$, we obtain 
\begin{eqnarray}
T(u\pi\varphi(x),\dots)&=&T_{\rm off}(\sigma\pi(u\varphi)(x),\dots)\nonumber\\
&=&\chi(u)_x\,T_{\rm off}(\varphi(x),\dots)\nonumber\\
&=& \chi(u)_x\,T(\pi\varphi(x),\dots)\ ,\label{T(basic)}
\end{eqnarray}
by using (\ref{T_on-T_off}), (\ref{sigma-chi}), the AWI and $\varphi=\sigma\pi(\varphi)$.
The same relation holds true for $u\pi\phi$ (complex scalar field), $u\pi\psi$ (Dirac field)
and $u\pi A^\mu_a$ (gauge field).

(3) As we see from the preceding Remark, $(\chi(u)-u)_x T(\pi A(x),\dots)$ 
gives directly the terms violating the AWI in the on-shell
formalism if $\sigma\pi(uA)=\chi(u)\, A$. However, the latter formula does generally not hold if 
$A\not\in\mathcal{P}_1$, e.g. $\sigma\pi(\partial_\mu\partial_\nu\,\varphi^2)\not=
\chi(\partial_\mu\partial_\nu)\,\varphi^2$. The terms violating the AWI can generally be expressed by
\begin{eqnarray}
\partial_\mu^x T(\pi A(x),\dots)-T(\partial_\mu\pi A(x),\dots)=
T_{\rm off}([\partial_\mu,\sigma\pi]\,A(x),\dots)\ .\label{[d,T]}
\end{eqnarray}
The appearing commutator of $\partial_\mu$ with $\sigma\pi$ can be expressed in terms of $\chi$
by using the Leibniz rule, more precisely $\partial_\mu A=\sum_a\frac{\partial A}{\partial (\partial^a\varphi)}\,
\partial_\mu\partial^a\varphi$ and $\partial_\mu \sigma\pi(A)=\sum_a
\frac{\partial\, \sigma\pi(A)}{\partial\,\sigma\pi(\partial^a\varphi)}\,
\partial_\mu\,\sigma\pi(\partial^a\varphi)$ (cf.~formula (108) in Ref.~\onlinecite{Dutsch03}):
\begin{eqnarray*}
[\partial_\mu,\sigma\pi]\,A=\sum_a\sigma\pi\Bigl(\frac{\partial A}{\partial (\partial^a\varphi)}\Bigr)\,
\Bigl(\partial_\mu\chi(\partial^a)-\chi(\partial_\mu\partial^a)\Bigr)\varphi\ .
\end{eqnarray*}

(4) The Master Ward Identity (MWI), which is a universal renormalization condition,
can be written as a formula for $T_{\rm off}((B\,[\partial_\mu,\sigma\pi]\,A)(x),\dots)\,\,(A,B\in \mathcal{P})$,
which is a generalization of (\ref{[d,T]}). (This is the version of the MWI 
given in formula (110) of Ref.~\onlinecite{Dutsch03}, which agrees essentially 
with the original formulation of the MWI\cite{Dutsch02} in terms of the 
on-shell formalism.) In this form, the MWI contains the differential operator
$\delta^\mu_{\partial^a\varphi,\partial^b\varphi}$ which can be written 
as\cite{chifn7}
\begin{eqnarray*}
\delta^\mu_{\partial^a\varphi,\partial^b\varphi}\,f(x)=-i\int dy\,T_{\rm off}\Bigl([\partial^\mu,\sigma\pi]\,
\partial^a\varphi(x),\partial^b\varphi(y)\Bigr)\,f(y)\ .
\end{eqnarray*}
To get a more explcit expression for $\delta^\mu_{\partial^a\varphi,\partial^b\varphi}$
we note that, due to (\ref{chi=}), there exists a map $\chi_1\,:\,S(V)\rightarrow S(V)$ such that
\begin{eqnarray*}
\chi(u)-u=\chi_1(u)\,(\Box +m^2)\ .
\end{eqnarray*}
With that and by using the AWI and
$(\Box+m^2)\Delta^F_m=-i\,\delta$ we obtain
\begin{eqnarray}
\delta^\mu_{\partial^a\varphi,\partial^b\varphi}=(-1)^{|b|+1}\,\partial^b\,
\Bigl(\partial^\mu\,\chi_1(\partial^a)-\chi_1(\partial^\mu\partial^a)\Bigr)\ .\label{delta^mu}
\end{eqnarray}
(Cf.~Appendix A of Ref.~\onlinecite{Dutsch02} where, in an unelegant way, 
examples for (the integral kernel of)
$\delta^\mu_{\partial^a\varphi,\partial^b\varphi}$ are computed in the on-shell
formalism without using the relation to the off-shell formalism 
(i.e.~the map $\sigma$).) 
If we know $\chi_1$ explicitly we can write down the MWI explicitly by 
using (\ref{delta^mu}).

(5) Condition (b) (or (b') respectively) can be obtained directly from (\ref{T(basic)}) and multilinearity:
$0=T(u(\Box+m^2)\,\pi\varphi(x),\dots)=\chi(u(\Box+m^2))_x\,T(\pi\varphi(x),\dots)$
implies (b), and similarly for (b').
\section{Construction of $\sigma$: scalar fields}

The map $\chi: S(V) \to S(V)$ is defined as follows:
for $u=\partial_{\mu_1}\dots\partial_{\mu_n}$,
$\chi(u)$ can be written as a sum over polynomials
$P_k^n(u)$ in $S(V)$
\begin{eqnarray}
\chi(u) &=&
\sum_{k=0}^{n/2}  \alpha_k^n P_k^n(u)\ ,
\label{expansion}
\end{eqnarray}
where $\alpha_k^n$ are Lorentz invariant coefficients to be determined
and $P_k^n(u)$ are specific covariant symmetric polynomials containing
$k$ metric tensors $g_{\lambda\mu}$ and
$(n-2k)$ partial derivatives. The upper index $n$ of $P^n_k$ refers to the
degree $|u|=n$ of $u$. We point out that the $\alpha_k^n$'s may contain 
partial derivatives in the form of powers of $\Box$ and they are functions of $m^2$.
To simplify the notation we write $n/2$ for the upper bound of the sum instead
of $\{n/2,(n-1)/2\}\cap {\bf N}_0$.

\subsection{Definition and properties of $P_k^n(u)$}
Let us start with a few examples and define
\begin{eqnarray*}
P_0^0(1) &=& 1,\\
P_0^1(\partial_\lambda) &=& \partial_\lambda,\\
P_0^2(\partial_\lambda\partial_\mu) &=& \partial_\lambda\partial_\mu,\\
P_1^2(\partial_\lambda\partial_\mu) &=& g_{\lambda\mu},\\
P_0^3(\partial_\lambda\partial_\mu\partial_\nu) &=& 
  \partial_\lambda\partial_\mu\partial_\nu,\\
P_1^3(\partial_\lambda\partial_\mu\partial_\nu) &=&
g_{\lambda\mu}\partial_\nu+
g_{\lambda\nu}\partial_\mu+
g_{\mu\nu}\partial_\lambda.
\end{eqnarray*}
It is clear that these
$P_k^n(u)$, for $k=0$ to $n/2$ are covariant and span
the space of symmetric Lorentz-covariant tensors of
rank $n$ that can be constructed from partial derivatives
and metric tensors without contracting any Lorentz indices,
i.e. $g_{\mu\nu}\,\partial^\mu\partial^\nu=\Box$ is excluded. 
A tensor with $n$ indices and $k$ metric tensors
that transforms under Lorentz transformation as
$u=\partial_{\mu_1}\dots\partial_{\mu_n}$
is
$u=g_{\mu_1\mu_2}\dots g_{\mu_{2k-1}\mu_{2k}}
\partial_{\mu_{2k+1}}\dots\partial_{\mu_{n}}$.
However, this tensor is not symmetric in its $n$ variables.
When we symmetrize it we obtain $P_k^n(u)$.
This is essentially the method used in Ref.~\onlinecite{Dutsch03}.

It will be convenient to be a little more formal and
to define a derivation $\delta^\mu$ on $S(V)$
by $\delta^\mu \partial_\nu = \delta^{\mu,\nu}$, where
$\delta^{\mu,\nu}$ is the Kronecker symbol equal
to 1 if $\mu=\nu$ and to 0 otherwise.
With this derivation, we define the operator
$\Lambda=(1/2) g_{\mu\nu} \delta^\mu\delta^\nu$
(Einstein summation convention is used throughout
for Lorentz indices).
The polynomials $ P_k^n(u)$ are defined by
\begin{eqnarray}
P_k^n(u) &=& \frac{1}{k!}\Lambda^k(u), \label{defPkn}
\end{eqnarray}
with $P_0^n(u)=u$.  Note that the linearity of $\Lambda$
as an operator acting on $S(V)$ 
implies the linearity of $P_k^n$:
$P_k^n(\lambda u + \mu v)=\lambda P_k^n(u) + \mu P_k^n(v)$.

We shall also use
\begin{eqnarray}
P(u)&\equiv & \exp(s\Lambda) u\,\quad
s\in {\bf R}\ , \label{exp}
\end{eqnarray}
so that $P(u) = \sum_{k=0}^{n/2} s^k P_k^n(u)$.
For notational convenience we use the shorthand notation $P(u)$
in which the dependence of $P$ on $s$ is left implicit.
Note that $\delta^\mu$ commutes with $\Lambda$ and, hence, it holds
\begin{eqnarray}
\delta^\mu P(u) &=& P(\delta^\mu u).
\label{partialmuP(u)}
\end{eqnarray}

Due to 
\begin{eqnarray}
\Lambda(\partial_{\mu_1}...\partial_{\mu_n}) &=&\sum_{i<j}
g_{\mu_i\mu_j}\,\partial_{\mu_1}...\hat i...\hat j...\partial_{\mu_n},\label{Lambda-explicit}
\end{eqnarray}
the polynomial $P_k^n(u)$ can be seen 
as the sum of all possible terms obtained
from $u=\partial_{\mu_1}\dots\partial_{\mu_n}$
by contracting $k$ pairs of indices,
where the contraction of $\partial_\lambda$
and $\partial_\nu$ is $g_{\lambda\nu}$.

From (\ref{Lambda-explicit}) we see that $\Lambda u$ is symmetric: $\Lambda(\tau u)=\Lambda(u)$
where $\tau u=\partial_{\mu_{\tau(1)}}\dots\partial_{\mu_{\tau(n)}}$ for any permutation
$\tau$ of $\{1,\dots,n\}$. It follows that also $P_k^n(u)$ is symmetric: $P^n_k(\tau u)=P^n_k(u)$
for all $\tau$.

The following property is useful to derive recursive proofs
\begin{eqnarray}
P(\partial_\mu u) &=& \partial_\mu P(u) + s g_{\mu\alpha}
P(\delta^\alpha u).\label{id1}
\end{eqnarray}
To prove this, we first use the definition of $\Lambda$
to get ${[}\Lambda,\partial_\mu{]}=g_{\mu\alpha} \delta^\alpha$.
Then, a recursive proof leads to
${[}\Lambda^k,\partial_\mu{]}=k g_{\mu\alpha} \delta^\alpha
\Lambda^{k-1}$ that yields 
$P(\partial_\mu u) = \partial_\mu P(u) + s g_{\mu\alpha}\delta^\alpha P(u)$.
Equation (\ref{id1}) follows from this and equation
(\ref{partialmuP(u)}).
If we repeat the same argument, we obtain
\begin{eqnarray}
P(\partial_\mu \partial_\nu u) &=& 
\big(\partial_\mu \partial_\nu + s g_{\mu\nu}\big) P(u) 
+ s g_{\mu\alpha}\partial_\nu P(\delta^\alpha u)
+ s g_{\nu\beta}\partial_\mu P(\delta^\beta u)
+ s^2 g_{\mu\alpha} g_{\nu\beta} P(\delta^\alpha \delta^\beta u).
\label{Pdmudnu}
\end{eqnarray}
In particular, if we consider $\Box=g^{\mu\nu} \partial_\mu
\partial_\nu$, the linearity of $P$ gives us
\begin{eqnarray*}
P(\Box u) &=& 
\big(\Box + s N\big) P(u) 
+ 2 s \partial_\mu P(\delta^\mu u)
+ s^2 g_{\mu\nu} P(\delta^\mu \delta^\nu u),
\end{eqnarray*}
where $N$ is the dimension of spacetime.

By equations (\ref{exp}), (\ref{partialmuP(u)}) 
the last term on the right hand side can be
rewritten as $2 s^2 \Lambda P(u)= 2 s^2 d P(u)/ds$.
Moreover, using (\ref{id1}) we get
\begin{eqnarray}
\partial_\mu P(\delta^\mu u) &=&
 P( \partial_\mu \delta^\mu u)
- s g_{\mu\alpha} \delta^\alpha \delta^\mu P(u)
\nonumber\\&=& |u| P(u) - 2 s \frac{d P(u)}{d s},
\label{partialdelta}
\end{eqnarray}
where 
we have used Euler's formula for homogeneous functions
to write $\partial_\mu \delta^\mu u=|u| u$.
Therefore
\begin{eqnarray*}
P(\Box u) &=& 
\big(\Box + s N + 2 s |u| - 2 s^2 d/ds\big) P(u).
\end{eqnarray*}
If we expand $P$ over $s$, this equality gives us
the crucial relation
\begin{eqnarray}
P^{n+2}_k(\Box u) &=& 
\Box P^n_k(u) + (N+2n-2k+2) P^n_{k-1}(u).
\label{Pnk(Boxu)}
\end{eqnarray}

\subsection{Determination of $\chi$}
From the ``on-shell condition'' (b) we will determine the coefficients $\alpha^n_k$
in the expansion (\ref{expansion}), using additionally
condition (e), which implies that
$\alpha_0^n=1$ for all $n\ge0$.
The on-shell condition becomes
\begin{eqnarray*}
\sum_{k=0}^{(n+2)/2} \alpha_k^{n+2} P_k^{n+2}(\Box u)+
m^2 \sum_{k=0}^{n/2} \alpha_k^n P_k^n(u) &=& 0.
\end{eqnarray*}
The relation (\ref{Pnk(Boxu)}) and the independence of
the polynomials $P_k^n$ give us the recurrence relation.
\begin{eqnarray*}
\Box \alpha_{k-1}^{n+1} + (N+2n-2k) \alpha_k^{n+1}
+ m^2  \alpha_{k-1}^{n-1} &=&0.
\end{eqnarray*}
This equation determines $\alpha_k^{n}$ in terms of $\alpha_0^{n}$.
Using the boundary condition $\alpha_0^{n}=1$ we obtain
$\alpha^n_1=-(\Box+m^2)/(N+2n-4)$. It can
be checked  straightforwardly that the general solution for $k>0$ is
\begin{eqnarray}
\alpha_k^n &=& (-1)^{k}\, (\Box+m^2)\,\sum_{p=0}^{k-1} {k-1 \choose p}
   m^{2p}\Box^{k-1-p}
  \prod_{q=0}^{k-1} (N+2 n - 2 p  - 2 q-4)^{-1}\ .
\label{expalphak}
\end{eqnarray}
This can also be written in terms of hypergeometric functions,
\begin{eqnarray*}
\alpha_k^n &=& \frac{(-1)^k m^{2k-2}}{2^{k}}
  \frac{\Gamma(N/2+n-2k)}{\Gamma(N/2+n-k)}\,\,
{}_2F_1(1-k,N/2+n-2k;N/2+n-k;-\Box/m^2)\,(\Box+m^2)\ .
\end{eqnarray*}
With this equation,
the many identities satisfied by hypergeometric functions can 
be used to obtain alternative expressions for $\alpha_k^n$.

We give now a few examples of $\chi(u)$ for low degrees:
\begin{eqnarray}
\chi(\partial_\lambda\partial_\nu) &=& 
\partial_\lambda\partial_\nu - \frac{g_{\lambda\nu}}{N} (\Box+m^2),\label{2derivatives}
\end{eqnarray}
which gives $\chi(\Box)= - m^2$ as it must be. 
Using additionally (\ref{T(basic)})
and $(\Box+m^2)\Delta^F_m=-i\,\delta$, it results
\begin{eqnarray*}
T(\partial_\lambda\partial_\nu\,\pi\varphi(x),\pi\varphi(y))&=&
\partial_\lambda^x\partial_\nu^x T(\pi\varphi(x),\pi\varphi(y))
+i\frac{g_{\lambda\nu}}{N}\,\delta(x-y)\ ,
\end{eqnarray*}
where it is used that the values of $T$ are on-shell.

In the case of a product of three derivatives we get
\begin{eqnarray*}
\chi(\partial_\lambda\partial_\mu\partial_\nu) &=& 
\partial_\lambda\partial_\mu\partial_\nu - 
\frac{\Box+m^2}{N+2} P^3_1(\partial_\lambda\partial_\mu\partial_\nu).
\end{eqnarray*}
\section{Construction of $\sigma$: Dirac fields}
If we consider the Dirac equation instead of the
Klein-Gordon equation, we can use the 
$\gamma$-matrices $\gamma_\mu$ to build
Lorentz covariant tensors. 
Since a symmetric Lorentz-covariant
tensor can contain at most one $\gamma$-matrix,
we should consider symmetric polynomials
containing zero and one $\gamma$-matrix.
The first ones are the $P^n_k(u)$ defined in the last
section, the second ones are called $Q^n_k(u)$
and defined by replacing in turn each $\partial_{\mu}$
of $P^n_k(u)$ by $\gamma_\mu$.

\subsection{Properties of $Q^n_k(u)$}
According to the previous discussion, the 
polynomials $Q^n_k(u)$ are defined by
$Q^n_k(u)= \gamma_\nu P^n_k(\delta^\nu u)$.
Let us give a few examples
\begin{eqnarray*}
Q_0^0 &=& 0,\\
Q_0^1(\partial_\lambda) &=& \gamma_\lambda,\\
Q_0^2(\partial_\lambda\partial_\mu) &=& \gamma_\lambda\partial_\mu
   +\partial_\lambda\gamma_\mu,\\
Q_1^2(\partial_\lambda\partial_\mu) &=& 0,\\
Q_0^3(\partial_\lambda\partial_\mu\partial_\nu) &=& 
  \gamma_\lambda\partial_\mu\partial_\nu+
  \partial_\lambda\gamma_\mu\partial_\nu+
  \partial_\lambda\partial_\mu\gamma_\nu,\\
Q_1^3(\partial_\lambda\partial_\mu\partial_\nu) &=&
g_{\lambda\mu}\gamma_\nu+
g_{\lambda\nu}\gamma_\mu+
g_{\mu\nu}\gamma_\lambda.
\end{eqnarray*}
Note that $Q_k^n=0$ if $k>(n-1)/2$.

As for the boson case, we need some recursive 
relations between these polynomials.
Defining 
\begin{eqnarray*}
Q(u)&:=&\gamma_\nu P(\delta^\nu u)
=\sum_k s^kQ^n_k(u)
\end{eqnarray*}
(as for $P$ the dependence on $s$ of $Q$ is suppressed in the notation),
we claim that
\begin{eqnarray}
Q(\partial_\mu u) &=& \partial_\mu Q(u) + \gamma_\mu P(u)
+s g_{\mu\alpha} Q(\delta^\alpha u).
\label{id2}
\end{eqnarray}
To show this, we start from equation (\ref{id1})
to obtain
\begin{eqnarray*}
Q(\partial_\mu u) &=& 
\gamma_\nu P(\delta^\nu\partial_\mu u)
=\gamma_\mu P(u) + \gamma_\nu P(\partial_\mu\delta^\nu u)
\\&=& \gamma_\mu P(u) +
\gamma_\nu \partial_\mu P(\delta^\nu u) + s \gamma_\nu 
g_{\mu\alpha}\delta^\alpha P(\delta^\nu u) \\
&=& \gamma_\mu P(u) +\partial_\mu Q(u) + s g_{\mu\alpha}\delta^\alpha
Q(u).
\end{eqnarray*}
In addition we deduce the second set of identities:
\begin{eqnarray}
\gamma^\mu P(\partial_\mu u) &=& P(u) (\gamma\cdot\partial)+ s Q(u),
\label{idgam1}\\
\gamma^\mu Q(\partial_\mu u) &=& (N+2|u|) P(u)
 -2 s \frac{d P(u)}{d s} - Q(u) (\gamma\cdot\partial).
\label{idgam2}
\end{eqnarray}
To prove the first identity (\ref{idgam1}) we start from
(\ref{id1}) and contract with $\gamma^\mu$:
\begin{eqnarray*}
\gamma^\mu P(\partial_\mu u) &=&  (\gamma\cdot\partial) P(u)+
s \gamma_\alpha\delta^\alpha P(u)
\\&=&  P(u) (\gamma\cdot\partial) + s Q(u),
\end{eqnarray*}
where we used the fact that the $\gamma$ matrices commute with
$P(u)$.
The second identity is a little bit more tricky.
We start from (\ref{id2}) and contract with $\gamma^\mu$:
\begin{eqnarray*}
\gamma^\mu Q(\partial_\mu u) 
&=& (\gamma\cdot \partial) Q(u) 
+ N P(u) +s \gamma_\mu \delta^\mu Q(u)
\\&=&
 (\gamma\cdot \partial) Q(u) 
+ N P(u) +s \gamma_\mu\gamma_\alpha  \delta^\mu \delta^\alpha P(u)
\\&=&
 (\gamma\cdot \partial) Q(u)
+ N P(u) + s g_{\mu\alpha} \delta^\mu \delta^\alpha P(u)
\\&=&
 (\gamma\cdot \partial) Q(u)
+ N P(u) + 2s \Lambda P(u),
\end{eqnarray*}
where we have used the fact that 
$\delta^\mu \delta^\alpha=\delta^\alpha\delta^\mu$ and
the definition of $\Lambda$.
Now, by equation (\ref{exp}), 
we have $\Lambda P(u) = d P(u)/ds$ and we obtain
\begin{eqnarray*}
\gamma^\mu Q(\partial_\mu u) &=& (\gamma\cdot \partial) Q(u)
+N P(u) +2 s \frac{d P(u)}{d s}.
\end{eqnarray*}
To obtain (\ref{idgam2}) we need to commute $\gamma\cdot \partial$
with $Q(u)$:
\begin{eqnarray*}
(\gamma\cdot \partial) Q(u) &=& \partial^\mu \gamma_\mu \gamma_\alpha 
\delta^\alpha P(u)
\\&=&
2\partial^\mu g_{\mu\alpha} \delta^\alpha P(u)
-  \partial^\mu \gamma_\alpha \delta^\alpha P(u) \gamma_\mu
\\&=&
2\partial_\mu \delta^\mu P(u)
-  Q(u) (\gamma\cdot \partial).
\end{eqnarray*}
Then, equation (\ref{partialdelta}) 
concludes the proof of equation (\ref{idgam2}).

From (\ref{idgam1}) and (\ref{idgam2}) we deduce identities
for the polynomials. For any 
 $u=\partial_{\nu_1}\dots\partial_{\nu_n}$ we have
\begin{eqnarray}
\gamma^\mu P_k^{n+1}(\partial_\mu u) &=& 
P_k^n(u) (\gamma\cdot\partial)+ Q_{k-1}^n(u)\ ,\label{idp}\\
\gamma^\mu Q_k^{n+1}(\partial_\mu u) &=& (N+2n-2k) P_k^n(u)
  - Q_k^n(u) (\gamma\cdot\partial)\ .
\label{idq}
\end{eqnarray}

\subsection{Determination of $\chi$}

To determine $\chi$, we use the fact that, by Lorentz covariance
and symmetry, the map $\chi$ can be written
\begin{eqnarray*}
\chi(u) &=&
\sum_{k=0}^{n/2} P_k^n(u)\,\alpha_k^n 
   +  \sum_{k=0}^{(n-1)/2} Q_k^n(u)\,\beta_k^n\ ,
\end{eqnarray*}
where $\alpha_k^n$ and $\beta_k^n$ are
Lorentz-invariant, they may contain $\Box$ and $(\gamma\cdot\partial)$
and depend on $m^2$. In particular for $n=0$ it results:
$1=\chi(1)=\alpha_0^0$.

The on-shell condition (b') imposes that
\begin{eqnarray*}
0 &=&
\sum_{k=0}^{(n+1)/2} i\gamma^\mu P_k^{n+1}(\partial_\mu u) 
      \alpha_k^{n+1}
 -m \sum_{k=0}^{n/2}  P_k^n(u) \alpha_k^n
   +  \sum_{k=0}^{n/2} i\gamma^\mu Q_k^{n+1}(\partial_\mu u) 
      \beta_k^{n+1} 
   -m  \sum_{k=0}^{(n-1)/2}  Q_k^n(u) \beta_k^n.
\end{eqnarray*}
Using the linear independence of all $ P_k^n$ and $Q_k^n$
and the identities (\ref{idp}) and (\ref{idq}),
the on-shell condition yields
the following system of equations
\begin{eqnarray}
i(\gamma\cdot\partial) \alpha_k^{n+1} + 
i (N+2n-2k) \beta_k^{n+1} &=& m \alpha_k^{n},\label{eq1}\\
i  \alpha_{k+1}^{n+1} -i (\gamma\cdot\partial) \beta_k^{n+1}
&=& m \beta_k^{n}.
\label{eq2}
\end{eqnarray}
The boundary condition imposed by (e') is
$\alpha_0^n=1$ for $n\ge0$. Equation (\ref{eq1})
implies that $\beta_0^n=i D/(N+2n-2)$ for $n>0$,
where $D$ is the Dirac operator, $D=i(\gamma\cdot\partial)-m$.

The solution of this system of equations is unique,
and can be given explicitly. By eliminating $\alpha_k^n$
from equations (\ref{eq1}) and (\ref{eq2}), we obtain
a recursive equation for $\beta_k^n$
\begin{eqnarray*}
\beta_k^{n+1} &=& -\frac{1}{N+2n-2k}
  \big(\Box \beta_{k-1}^{n+1}+m^2  \beta_{k-1}^{n-1}\big),
\end{eqnarray*}
with the boundary condition $\beta_0^n=i D/(N+2n-2)$.
The solution of this equation is
\begin{eqnarray*}
\beta_k^n &=& i(-1)^{k}\,D\,\sum_{p=0}^k {k \choose p} m^{2p}\Box^{k-p}
  \prod_{q=0}^k (N+2 n - 2 p  - 2 q-2)^{-1}\ ,
\end{eqnarray*}
where $k\geq 0$ and $n\geq 2k+1$.
This sum is proportional to a hypergeometric function:
\begin{eqnarray*}
\beta_k^n &=& i\frac{(-m^2)^{k}}{2^{k+1}}
  \frac{\Gamma(N/2+n-2k-1)}{\Gamma(N/2+n-k)}\,\,
{}_2F_1(-k,N/2+n-2k-1;N/2+n-k;-\Box/m^2)\,D\ .
\end{eqnarray*}

The explicit expression for $\alpha_k^n$ (for $k\geq 1\,,\,n\geq 2k\geq 2$)
is then obtained from
(\ref{eq2}). Since $\beta_k^n\in S(V)\, D\,\,,\,\forall n,k$
and $\alpha_k^n\in (S(V)+S(V)\,(\gamma\cdot\partial))\,D\,\,,\,\forall k\geq 1,\, n\geq 2$
we obtain $(\chi(u)-u)\in S(V)^{f(N)\times f(N)}\, D$ and, hence, 
$(\chi(u)-u)\psi\in\mathcal{J}\,,\,\forall u\in S(V)$.
 
Examples:
\begin{eqnarray*}
\chi(\partial_\mu) &=& \partial_\mu + \frac{i}{N}\gamma_\mu\, D,\\
\chi(\partial_\mu\partial_\nu) &=& 
   \partial_\mu \partial_\nu + \frac{i}{N+2}  
   (\gamma_\mu \partial_\nu + \partial_\mu \gamma_\nu)\, D 
  + \Big(\frac{i\gamma\cdot\partial}{N+2} 
   + \frac{m}{N}\Big)\, g_{\mu\nu}\, D.
\end{eqnarray*}
These results can be checked by verifying the on-shell condition, e.g.~the
first formula fulfills indeed $\gamma^\mu\,\chi(\partial_\mu)=-im$.
\section{Construction of $\sigma$: gauge fields}
To complete the determination of $\sigma$, we now consider the case of gauge
fields. In contrast to the previous cases, the map $\sigma$
is not unique. Still, the complete solution is determined and
uniqueness can be recovered by using a simplifying assumption.

\subsection{Conditions on $\chi$}
We study gauge fields $(A^\mu_a)_{a=1,...,M}$ in Feynman gauge without 
introducing the auxiliary fields $B_a$ (the Nakanishi-Lautrup fields),
where $a$ is the 'colour index'. 
With that the free field equations read
\begin{eqnarray}
\square A^\mu_a&=&0\ ,\quad a=1,...,M\ .
\end{eqnarray}
Notice also that gauge fields are real: $A^{\mu\,\star}_a= A^\mu_a$.
The commutative field algebra $\mathcal{P}$ and the ideal 
     $\mathcal{J}$ of the free field equation read
\begin{eqnarray*} 
\mathcal{P}:= \bigvee\{\partial^a A_{\mu ,b}\,|\,\mu=0,...,N-1;\,b=1,...,M;\,
a\in{\bf N}_0^N\}
\end{eqnarray*} 
and
\begin{eqnarray*} 
        \mathcal{J}=\{\sum_{a,\mu,b} B_{a,\mu;b}\,\partial^{a} 
        \square A^\mu_b\,|\, B_{a,\mu;b}\in\mathcal{P}\} \ ,
\end{eqnarray*} 
respectively.

There is no restriction on $\sigma$ coming from gauge invariance 
(or BRST-invariance) for the following reason (cf.~Sect.~5.2 in Ref.~\onlinecite{Dutsch03}): 
$\sigma$ appears only in the {\it entries} of the time-ordered product 
$T_\mathrm{off}$ (\ref{T_on-T_off}), however, gauge invariance concerns the 
{\it values} (usually the on-shell values) of the time-ordered product.
The map $\sigma$ is determined by the axioms (i)-(iv) for one real scalar 
field and the following requirements:
\begin{enumerate}
     \item [(v'')] Since gauge fields are massless, $\sigma\pi$ must 
         {\it maintain} the mass dimension.
     \item [(vi'')] $\sigma\pi$ is diagonal in the colour index $a$. 
            Hence, the latter is omitted in the following.
\end{enumerate}
It immediately follows that 
\begin{eqnarray}
\sigma\pi(A_\nu)&=&A_\nu\ ,\quad \sigma\pi(\partial_\mu A_\nu)=
\partial_\mu A_\nu\ .
\end{eqnarray}
Again, axiom (iv) is equivalent to the existence of a map 
    $\chi:S(V)\rightarrow S(V)^{N\times N}$ such that
\begin{eqnarray}
\sigma\pi(u\,A_\nu)&=&\chi(u)_\nu^{\,\,\tau} A_\tau\ ,\quad\forall u\in S(V)\ .
\end{eqnarray}
It follows that $\chi$ must satisfy the properties (a), (b) (with $m=0$), 
   (c) listed in Sect.~5 and
\begin{enumerate}
     \item [(d'')] $\chi$ maintains the degree of the derivatives, i.e.
             $\chi(\partial_{\mu_1}...\partial_{\mu_n})_\nu^{\,\,\tau}$ is a 
          linear combination of derivatives of order $n$.
     \item [(e'')] On monomials $u=\partial_{\mu_1}\dots\partial_{\mu_n}$, 
        $\chi$ takes the form
\begin{eqnarray}
\chi(u)_{\nu\tau} &=& \sum_{k=0}^{n/2} a^n_k P^n_k(u) g_{\nu\tau}
  +\sum_{k=0}^{(n-2)/2} b^n_k \sum_{\alpha\beta} g_{\nu\alpha} g_{\tau\beta}
    P^{n-2}_k(\delta^\alpha \delta^\beta u)
  +\sum_{k=1}^{(n-1)/2} c^n_k \sum_{\alpha} g_{\nu\alpha} 
          P^{n-1}_k(\delta^\alpha u) \partial_\tau
\nonumber\\&&
  +\sum_{k=1}^{(n-1)/2} d^n_k \sum_{\alpha} g_{\tau\alpha} 
       P^{n-1}_k(\delta^\alpha u) \partial_\nu
  + \sum_{k=2}^{n/2} e^n_k P^n_k(u) \partial_\nu  \partial_\tau
\label{chinutau}
\end{eqnarray}
with $a_0^n=1$.
\end{enumerate}
To prove (e''), we proceed analogously to \eqref{chi=}, and we
notice that the indices $\nu$ and $\tau$ can appear either in
a partial derivative or in a metric tensor. If they both appear
in a derivative we get the last term, if only one of them
appears in a derivative we get the third and fourth terms.
If they both appear in a metric tensor, they can be in the
same tensor (this gives the first term) or in two different
tensors (as in the second term).
Note that all these terms appear in the right hand side
of equation \eqref{Pdmudnu}. Condition (d'')
implies that $a^n_k$, $c^n_k$ and $d^n_k$ are of degree $k$ in $\Box$,
$b^n_k$ is of degree $k+1$ in $\Box$ and
$e^n_k$ is of degree $k-1$ in $\Box$. $a_0^n$ and the
range of values of $k$ are determined by the fact that
the only term without $\Box$ must be equal to $u\,g_{\nu\tau}$.

The conditions (a),(b),(c), (d'') and (e'') on $\chi$ are also 
{\it sufficient} for (i)-(iv) and (v''): in contrast to the preceding cases 
this is obvious also for (i), because (e'') and the degree in $\Box$
of the coefficients directly imply that 
$\chi(u)-u\,\mathbf{1}=D(u)\square$ for some $D(u)\in S(V)^{N\times N}$.

\subsection{Determination of $\chi$}

The unknown coefficients in (e'') are restricted only by (b):
\begin{eqnarray}
g^{\mu_1\mu_2}\,\chi(\partial_{\mu_1}...\partial_{\mu_n})_{\nu\tau}=0\ .
\label{on-shell-restr}
\end{eqnarray}
This condition can be rewritten
$\chi(\Box v)=0$, with
$v=\partial_{\mu_3}\dots\partial_{\mu_n}$.
To determine the coefficients in equation \eqref{chinutau}
for $u=\Box v$, we need the  following identities
\begin{eqnarray}
 g_{\nu\alpha} \delta^\alpha \Box v &=& 
 2 \partial_\nu v
 + g_{\nu\alpha} \Box \delta^\alpha v,
 \label{gna}
 \end{eqnarray}
and
\begin{eqnarray}
 g_{\nu\alpha} g_{\tau\beta} \delta^\alpha \delta^\beta \Box v &=& 
  2 g_{\nu\tau} v
 + 2 g_{\tau\alpha} \partial_\nu \delta^\alpha v
 + 2 g_{\nu\alpha} \partial_\tau \delta^\alpha v
 + g_{\nu\alpha} g_{\tau\beta} \Box \delta^\alpha \delta^\beta v.
 \label{gnab}
\end{eqnarray}
Now we use equation \eqref{Pnk(Boxu)} to calculate
\begin{eqnarray*}
 P^{n}_k(\Box v) &=& 
 \Box P^{n-2}_k(v) + (N+2n-2k-2) P^{n-2}_{k-1}(v).
 \end{eqnarray*}
Equation \eqref{id1} gives us
 \begin{eqnarray}
 P^{n-1}_k(\partial_\nu v) &=& P^{n-2}_k(v)\partial_\nu
 + g_{\nu\alpha} P^{n-3}_{k-1}(\delta^\alpha v).
 \label{Pn-1}
 \end{eqnarray}
This, together with equations \eqref{gna} and \eqref{Pnk(Boxu)}, yields
 \begin{eqnarray*}
 g_{\nu\alpha} P^{n-1}_k(\delta^\alpha\Box v) &=& 
 2 P^{n-2}_k(v) \partial_\nu + 
 \Box g_{\nu\alpha} P^{n-3}_{k}(\delta^\alpha v)
 + (N+2n-2k-2) g_{\nu\alpha} P^{n-3}_{k-1}(\delta^\alpha v).
 \end{eqnarray*}
 Similarly, equations \eqref{gnab}, \eqref{Pn-1} and \eqref{Pnk(Boxu)} give us
 \begin{eqnarray*}
 g_{\nu\alpha} g_{\tau\beta} P^{n-2}_k(\delta^\alpha \delta^\beta \Box v) &=& 
 2 P^{n-2}_k(v) g_{\nu\tau} 
 + 2g_{\tau\alpha} P^{n-3}_k(\delta^\alpha v) \partial_\nu
 + 2g_{\nu\alpha} P^{n-3}_k(\delta^\alpha v) \partial_\tau
 + g_{\nu\alpha} g_{\tau\beta} \Box P^{n-4}_k(\delta^\alpha \delta^\beta v)
 \\&& + g_{\nu\alpha} g_{\tau\beta} (N+2n-2k-2) 
    P^{n-4}_{k-1}(\delta^\alpha \delta^\beta v).
 \end{eqnarray*}

If we gather all these results in equation \eqref{chinutau}, 
$\chi(\Box v)$ becomes
\begin{eqnarray*}
\chi(\Box v)_{\nu\tau} &=& \sum_k \big(\Box a^n_k +2 b^n_k
         +(N+2n-2k-4) a^n_{k+1}\big)\, g_{\nu\tau} P^{n-2}_{k}(v)
\\&&
   + \big(\Box e^n_k +2 c^n_k + 2 d^n_k
         +(N+2n-2k-4) e^n_{k+1}\big)\, \partial_\nu\partial_\tau P^{n-2}_{k}(v)
\\&&
   + \big(\Box d^n_k +2 b^n_k
        +(N+2n-2k-4) d^n_{k+1}\big)\, \partial_\nu g_{\tau\alpha}
            P^{n-3}_{k}(\delta^\alpha v)
\\&&
  + \big(\Box c^n_k +2 b^n_k
       +(N+2n-2k-4) c^n_{k+1}\big)\, \partial_\tau g_{\nu\alpha}
           P^{n-3}_{k}(\delta^\alpha v)
\\&&
   + \big(\Box b^n_k 
        +(N+2n-2k-4) b^n_{k+1}\big)\, g_{\nu\alpha} g_{\tau\beta} 
             P^{n-4}_{k}(\delta^\alpha\delta^\beta v)\ .
\end{eqnarray*}
The requirement $\chi(\Box v)_{\nu\tau}=0$ is equivalent to the condition that the coefficients
of the linearly independent tensors $g_{\nu\tau} P^{n-2}_{k}(v)$, $\partial_\nu\partial_\tau P^{n-2}_{k}(v)$,
$\partial_\nu g_{\tau\alpha} P^{n-3}_{k}(\delta^\alpha v)$, $\partial_\tau g_{\nu\alpha} P^{n-3}_{k}(\delta^\alpha v)$
and $g_{\nu\alpha} g_{\tau\beta} P^{n-4}_{k}(\delta^\alpha\delta^\beta v)$, respectively, vanish for all $k$.
These equations can be solved recursively. Their solutions can be written in terms of the real numbers
$F^n_k$ defined by $F^n_0=1$ and
\begin{eqnarray*} 
F^n_k=(-1)^k\prod_{j=1}^k (N+2n-2j-2)^{-1}
\end{eqnarray*} 
for $1\le k \le n/2$. It results
\begin{eqnarray*}
a^n_k &=& (2k \beta_n+1) F^n_k \Box^k,\\
b^n_k &=& \beta_n F^n_k \Box^{k+1},\\
c^n_k &=& d^n_k = 2k\beta_n F^n_k \Box^k,\\
e^n_k  &=& 4k(k-1)\beta_n F^n_k \Box^{k-1}.
\end{eqnarray*}
The real numbers $\beta_n$ are free parameters of the solution;
they can be interpreted as a parametrization of the unknown boundary value
$b^n_0$: $b^n_0=\beta_n \Box$.

The simplest non-trivial examples are
\begin{eqnarray*}
\chi(\partial_{\mu_1}\partial_{\mu_2})_{\nu\tau} &=&
\partial_{\mu_1}\partial_{\mu_2} g_{\nu\tau}
-\frac{1+2\beta_2}{N} \Box g_{\mu_1\mu_2} g_{\nu\tau}
+\beta_2 \Box (g_{\nu\mu_1} g_{\tau\mu_2} +g_{\nu\mu_2} g_{\tau\mu_1}).
\end{eqnarray*}
\begin{eqnarray*}
\chi(\partial_{\mu_1}\partial_{\mu_2}\partial_{\mu_3})_{\nu\tau} &=&
\partial_{\mu_1}\partial_{\mu_2}\partial_{\mu_3} g_{\nu\tau}
-\frac{1+2\beta_3}{N+2} \Box 
     (g_{\mu_1\mu_2}\partial_{\mu_3}+ \mathrm{c.p.}) g_{\nu\tau}
+\beta_3 \Box 
   (g_{\nu\mu_1} g_{\tau\mu_2}\partial_{\mu_3} 
    +g_{\nu\mu_2} g_{\tau\mu_1}\partial_{\mu_3} + \mathrm{c.p.})
\\&&
-\frac{2\beta_3}{N+2} \Box 
   (g_{\nu\mu_1}g_{\mu_1\mu_2}+ \mathrm{c.p.}) \partial_\tau
-\frac{2\beta_3}{N+2} \Box 
   (g_{\tau\mu_1}g_{\mu_1\mu_2}+ \mathrm{c.p.}) \partial_\nu,
\end{eqnarray*}
where $\mathrm{c.p.}$ denotes the two non-trivial cyclic permutations
of $\mu_1, \mu_2, \mu_3$.

The presence of the infinite number of parameters $\beta_n$ can be avoided by adding 
the following constraint
\begin{enumerate}
     \item [(vii'')] $\sigma\pi$ is diagonal in the coordinate index
            $A_\nu$ of $A$.
\end{enumerate}
In other words, $\chi(u)_{\nu\tau}$ is proportional
to $g_{\nu\tau}$. The only term in equation \eqref{chinutau}
that satisfies this condition is the first one, so that $\beta_n=0$ and
\begin{eqnarray*}
\chi(u)_{\nu\tau} &=& \sum_{k=0}^{n/2} F^n_k \Box^k P^n_k(u) g_{\nu\tau}.
\end{eqnarray*}
This solution for $a^n_k$ is the same as for 
the real scalar field with $m=0$;
the components $A_\nu$ of $A$ are treated 
as \emph{independent} scalar fields.
We point out that, differently to the conditions (i)-(vi'') on $\sigma$, 
(vii'') is not needed to relate correctly on-shell and off-shell formalism 
by (\ref{T_on-T_off}), it is motivated only by simplicity. 

\section{Concluding Remarks}
In this paper, we describe explicitly the map $\chi$ 
for the scalar, Dirac and gauge fields.
An explicit knowledge of this map is very useful for various purposes: 
\begin{itemize}
\item first of all to compute time ordered products of fields involving higher 
derivatives in the on-shell formalism. Knowing $\chi$, it is possible to go
to the off-shell formalism (\ref{T_on-T_off}), in which the computation is strongly 
simplified due to the validity of the AWI (an example is given in Remark (2)). In particular, using $\chi$,
the terms violating the AWI in the on-shell formalism can be expressed by a better computable 
time ordered product $T_\mathrm{off}(...)$ (Remark (3)).
\item to write down explicitly the 
Master Ward Identity, which has many important and far reaching applications
\cite{Dutsch02,Dutsch03,Brennecke} (see Remark (4)).
\end{itemize}  
The investigation of quantum field theories involving
high-order derivatives was not extensive in the literature
because they are generally not renormalizable. 
However, the renewed activity in nonrenormalizable quantum field theories
\cite{Gomis} makes them interesting again.
The terms with derivatives may be present already in the original
interaction Lagrangian
(i.e.~in the term of first order in the coupling constant with which
one starts the construction 
of the perturbation series, see e.g.~the treatment of perturbative
quantum gravity in Ref.~\onlinecite{Scharf2})
or they may enter the interaction only via the counter terms (explicit
examples are given in Ref.~\onlinecite{Kazakov}).

Even for {\it scalar} models it may happen that $\sigma$ is non-unique, e.g.~if one introduces
auxiliary fields which substitute for derivated fields, an example is worked out 
in Ref.~\onlinecite{Dutsch03}.

It would be useful to extend the present work to 
Dirac fields that are the solution
of the Dirac equation in the presence of an external potential. 

\medskip
\noindent{\bf Acknowledgment:}
We profitted from discussions with Klaus Fredenhagen and Raymond Stora.
We are grateful also to Bertfried Fauser for helping us with the Clifford
algebraic aspects of this work.




\begin{thebibliography}{40}
\bibitem[{chi({\natexlab{b}})}]{chifn2}
\bibinfo{note}{One may argue that the MWI is a too general renormalization
  condition since it cannot always be satisfied and that only certain cases of
  the MWI are needed for a consistent QFT. But, for example, a complete
  treatment of BRS-symmetry by means of the MWI is impossible within the usual
  on-shell formalism, see Example 4.1~in Ref.~\onlinecite{Dutsch02}.}
\bibitem[{chi({\natexlab{c}})}]{chifn3}
\bibinfo{note}{In Ref.~\onlinecite{Dutsch03} an additional axiom (vi) is given
  which requires that $\mathcal{P}$ is generated by fields in the image of
  $\sigma$ and their derivatives. This condition is needed to relate correctly
  the improved on-shell formalism of Ref.~\onlinecite{Dutsch02} to the
  off-shell formalism. Axiom (vi) is automatically satisfied if $\sigma\pi$
  maps each (underivated) basic field into itself, e.g. $\sigma\pi(\phi)=\phi$
  and $\sigma\pi(\phi^*)=\phi^*$ for the model of one complex scalar field
  $\phi$. The latter follows from axioms (i)-(v) for the models treated in this
  paper. But (vi) is non-trivial if one introduces auxiliary fields, which is
  quite often done to eliminate derivative couplings \cite{Dutsch03}.}
\bibitem[{chi({\natexlab{g}})}]{chifn7}
\bibinfo{note}{This is formula (111) in Ref.~\onlinecite{Dutsch03} translated
  into time ordered products. We use here the convention that the values of
  $T_\mathrm{off}$ are on-shell.}
\end{thebibliography}

\end{document}